\documentstyle[epsfig,aps]{revtex}

\def\be{\begin{eqnarray}}
\def\ee{\end{eqnarray}}
\def\nn{\nonumber\\ }
\def\Tr{{\rm Tr}\;}

\def\la{\langle}
\def\ra{\rangle}

\def\ie{ {\em i.e.} }
\def\eg{ {\em e.g.} }

\def\F{{\rm\bf F}}

\def\bH{{\rm\bf H}}

\def\J{{\rm\bf J}}

\def\R{{\rm\bf R}}

\def\W{{\rm\bf W}}

\def\bi{\begin{itemize}}
\def\ei{\end{itemize}}

\long\def\beginomit#1\endomit{}

\begin{document}
\title{Thermal Dileptons from a Nonperturbative Quark-Gluon Phase}

\author{C.-H. Lee$^a$, J. Wirstam$^b$,  I. Zahed$^a$ and T.H.
Hansson$^b$}
\address{a) Department of Physics \& Astronomy, SUNY at Stony Brook, 
Stony Brook, NY 11794, USA\\
b) Institute of Theoretical Physics, University of Stockholm,\\
Box 6730, S-113 85 Stockholm, Sweden}

\maketitle

\begin{abstract}
Assuming that gluon condensates are important even above the 
deconfining phase transition, we develop a model for the dilepton yield 
from a quark gluon plasma. Using a simple fire ball description of a  
heavy ion collision, and various estimates of the strengths of the gluon 
condensates, we compare our predicted dilepton yields with those
observed in the CERES and HELIOS experiments at CERN. The simple 
model gives an adequate description of the data, and in particular it 
explains the observed considerable enhancement of the yield in the low 
mass region. 

\end{abstract}

\section{Introduction and Summary}

Extensive lattice simulations have shown that QCD exhibits a phase 
transition at a temperature of the order of the infrared scale $\Lambda$.
This transition is believed to be a confinement-deconfinement transition 
for heavy quarks, and a chiral transition for light quarks. The high 
temperature phase is characterized by a small coupling constant $g(T)$.
Unfortunately, perturbation theory in $g(T)$ is plagued with infrared
singularities since the magnetic sector is non-perturbative at all
temperatures.

High temperature lattice~\cite{LAT1,LAT2} and instanton simulations~\cite{INST}
seem to concur on the fact that in the vicinity of the critical temperature, 
$T_{c}$, the chromoelectric and chromomagnetic condensates are still 
substantial. A number of theoretical arguments, both 
perturbative~\cite{AAP} and nonperturbative~\cite{AANP}, suggests that 
in the Euclidean formulation, the high temperature phase of QCD exhibits 
strong $A_4A_4$ correlations. 
It has even been argued that the global color symmetry might be
spontaneously 
broken with the appearance of an $A_4$ condensate~\cite{AANP,NADK}.
Also,  lattice simulations show that chiral symmetry is restored 
at the critical temperature, so that the quark condensate can be set to 
zero above $T_{c}$. 

In the present analysis we will incorporate these three important
observations that are pertinent to QCD  near and above $T_{c}$; 
the quark condensate vanishes, the chromoelectric and 
chromomagnetic correlations remain non-zero, and strong correlations 
develop in the static part of the $A_4$ gauge field that can be 
described by a local $A_{4}^{2}$ condensate. In the spirit of the QCD
sum rule method, we will not address the   complicated dynamical questions of 
how the gluon condensates come about, but simply assume that they exist.  
We will also restrict our discussion to only those with 
dimension 2, $\la \frac{\alpha_s}{\pi} A_4^2\ra$, and 4, $\la \frac{\alpha_s}{\pi} E^2\ra$ and $\la
\frac{\alpha_s}{\pi} B^2\ra$, 
following an earlier suggestion by two of us \cite{HZ90}. The value of 
the condensates can only be estimated using non-perturbative calculations such 
as lattice QCD, and at high temperature, the  scale is expected to be  set by 
the mass gap in the magnetic sector which is $\sim g^{2}T$.

That chromoelectric and chromomagnetic condensates could be present 
in the high temperature phase is not very surprising, and their size 
can be  roughly estimated using the bulk energy
and momentum measurements at finite temperature on the lattice. The 
analysis in~\cite{LAT2} gives the values : $\la \frac{\alpha_s}{\pi}E^2\ra\sim 
\la \frac{\alpha_s}{\pi}B^2\ra \sim (200 \; {\rm MeV})^4$ 

The presence of an $A^{2}_{4}$ condensate is more controversial~\cite{KAJA}, 
so we will give a  summary of the arguments presented in~\cite{HZ90} 
for why such a condensate should be considered. 
The basic observation is that the Wilson line (or Polyakov loop) 
operator $W={\rm Tr P}{\rm exp}({i\int^{\beta}_{0}  dx_{4}\, A_{4}(x_{4})  })$,
is a gauge invariant operator that is a good order parameter for the pure 
glue theory. Although this operator is non-local, it is possible to 
relate it to a string of local operators by choosing the static gauge 
$\partial_{4}A_{4}=0$. For example, in an SU(2) theory we have
$\langle W \rangle = \langle  1 +\cos(\frac \beta 2 A_{4})\rangle $, and one 
could now try to use factorization to extract $\la A^{2}_{4} \ra$. Here 
we will simply equate the condensate to the magnetic mass, \ie we 
will  typically use,  $\la \frac{\alpha_s}{\pi} A_4^2\ra/T^2
\sim\frac {\alpha_{s}} \pi (g^{2})^{2}\sim 16\pi\alpha_s^3$, 
which is about $0.4$ for $\alpha_s\sim 0.2$
This estimate is also supported by a lattice calculation in pure 
Yang-Mills \cite{AANP}. Indeed, an analysis of the gluon-propagator 
in four-dimensions in Landau-gauge gives $a^2\la (g A_4)^2\ra \sim 1/3$ on a 
$4\times 8^3$ lattice, which is equivalent to $\la \frac{\alpha_s}{\pi} 
A_4^2\ra/T^2 \sim 0.4$. A direct reconstruction of this condensate from the
dimensionally reduced theory has proven to be more subtle~\cite{KAJA}. 

In this letter, we present a calculation that suggests that the gluon
condensates strongly affect the emission rates of low mass dileptons 
from the plasma. Our calculation will be very naive. We will simply 
assume that the quark-antiquark annihilation processes that give rise 
to the dileptons occur in the presence of low momentum  fluctuating 
background gluon fields. By averaging over the background fields, we 
get leading corrections to the emission rates proportional to the 
various condensates. Note that in doing so we take the concept of 
a fluctuating background field at face value and perform the 
calculations directly in the time-like region assuming that the 
temperature is high enough for perturbation theory to be 
useful. Thus our method is very different from the QCD sum rules, 
where the background field method is just a convenient tool to calculate 
coefficient functions in the operator product expansion which is then 
used in the deep Euclidean region.\footnote{ A work  similar in spirit 
to ours is the calculation by Patkos and Sakai~\cite{PATKOS}
of the nonperturbative magnetic contribution to the $e^+e^-$
annihilation rate. That calculation was inspired by the so called Copenhagen 
vacuum based on fluctuating magnetic fluxtubes.         }
However, since the emission rates can be expressed in terms of 
current-current correlation functions we can use the standard QCD sum 
rule \emph{techniques} to calculate the condensate contributions. How 
this is done in detail, especially when it comes to gauge choice, is 
explained in \cite{HZ90}. To calculate the contribution from the 
$A^{2}_{4}$ condensate we will use the following observation: A 
constant potential $A_{4}$ is the same as an imaginary 
(color) chemical potential, $\mu^{a}$, and thus one can 
obtain the term $\sim \langle A^{2}_{4}\rangle$  by using a grand
canonical ensemble for the quarks with a chemical potential 
$\mu^{a} = iA^{a}_{4}$, \emph{averaged} over the arbitrary directions,
$a$, in color space, and pick up the term $\sim \mu^{2}$.  (We have also 
checked that this procedure gives the same result as the  
standard fixed point gauge methods used in \cite{HZ90}.)

At this point we want to stress that although our approach is 
inspired by the above theoretical arguments, it is basically 
phenomenological, and should primarily be judged by how successfully 
it can accommodate data. The theoretical uncertainties are not limited 
to the assumptions we made about the non-perturbative corrections to 
correlations functions in a high temperature quark gluon plasma. We 
also have to make several assumptions about the physics of the heavy 
ion collision, and we will use an oversimplified model, namely that 
of an expanding fireball containing {\it only} the plasma phase. The main
motivation for using such a simple model (the details are given in 
section III below) is to demonstrate that the gluon condensate effects 
are important. Hopefully more realistic models \eg including regions with 
mixed or hadronic phases, would also exhibit this sensitivity to the 
condensates. Also in the transition regime, the pure and mixed
quark condensates may still be relevant.

Having said all this, we will demonstrate that a naive fire ball model
combined with a  simple parametrization of the production amplitudes 
based on the above ideas, and with the condensate parameters at the 
values suggested by the scale arguments, is quite successful 
in explaining the dilepton data from the CERES and HELIOS 
relativistic heavy ion experiments. The rest of this paper, where this 
claim is substantiated, is organized as follows: In section~\ref{sec:emirate}, 
we calculate the effects of the leading order condensates on the dilepton 
emission rates from a thermalized quark-gluon phase. In 
section~\ref{sec:results}, we use a simple fire-ball 
model to estimate the  dilepton yield, and compare to the CERES and
HELIOS experiments. In section~\ref{sec:ptcut}, we show that our emission 
rates are also consistent with the recent analysis of the CERES 
results using different $p_t$ cuts.

\section{Emission Rates}
\label{sec:emirate}

In a translationally invariant, but non-perturbative, quark-gluon phase, 
the rate ${\bf R}$ of dileptons produced in a unit four volume is
directly related to the electromagnetic current-current correlation 
function~\cite{LARRY,WELDON}. For a pair of leptons with mass $m_l$ and 
momenta $p_{1,2}$, the rate per unit invariant momentum $q=p_1+p_2$ is ,
     \be
     \frac{d \R}{d^4 q} &=&-\frac{\alpha^2}{6\pi^3 q^2}
     \left(1+\frac{2 m_l^2}{q^2}\right)
     \left(1-\frac{4 m_l^2}{q^2}\right)^{1/2} \W (q) \ \ \  ,
     \ee
where $\alpha=e^2/4\pi$ is the fine structure constant, and
     \be
     \W (q) = \int d^4 x e^{-iq\cdot x} \Tr 
        \left({\bf P}e^{iA_4^a{\bf Q}^a} 
             e^{-(\bH-\F)/T} \J^\mu(x) \J_\mu(0)\right) \ \ \ .
     \label{eq:2}
     \ee
Here $e\J_\mu$ is the hadronic part of the electromagnetic current,
${\bf H}$  the hadronic Hamiltonian, $\F$  the free energy, 
$T$  the temperature, and ${\bf Q}^a$  a global color 
generator. (As mentioned above the direction, $a$, is arbitrary but 
fixed.) ${\bf W}(q)$ is related to the retarded current-current correlation 
function by,
   \be
   {\bf W} (q) = \frac{2}{e^{q^0/T}-1}\,{\rm Im}\,\W^R (q) \ \ \, 
   \label{eq2a}
   \ee
where
     \be
     \W^R(q) = i\int d^4 x e^{iq\cdot x}\Tr
        \left({\bf P}e^{iA_4^3 {\bf Q}^3}
        e^{-(\bH-\F)/T} {\bf R}^\star \left(\J^\mu(x) 
        \J_\mu(0)\right)\right) \ \ \ \ \ ,
\label{eq2b}
     \ee
and $ {\bf R}^\star$ is the retarded product. The leading contribution 
to the retarded function is condensate independent, and is just the 
``Born'' $q{\overline q}$ annihilation term, which for massless quarks 
becomes,
   \be
   {\rm Im} \W^R &=& \frac{1}{4\pi} \left( N_c\sum_{q=u,d,s}
e_q^2\right)
        q^2 \left[ 1+\frac{2T}{|\vec q|} 
    \ln\left(\frac{n_+}{n_-}\right)\right]
  \ \ \ ,  \label{eq:WR}
   \ee
where  $N_c$ is the number of colors, $e_q$ the charge 
of the quarks, and $n_\pm = 1/(e^{(q_0\pm |\vec q|)/2T}+1)$.
At $T=0$ this is just the $R_{\ge 3}$ ratio well known from 
$e^+e^-$ annihilation
   \be
   R_{\ge 3}(q^2) =\sum_i \frac{\sigma_i(\ge 3\pi)}
                          {\sigma(e^+e^-\rightarrow\mu^+\mu^-)}
                    = -\frac{4\pi}{q^2}\left. {\rm Im}\W^R_0
\right|_{T=0}
   = {N_c}\sum_{q} e_q^2 \ \ \  .
   \label{eq:R3}
   \ee
Using only the three light flavors, $R_{\ge 3}$ is  $2$.

As mentioned above, the contribution of the $A_4A_4$ condensate to the
retarded function
${\bf W}^R$ can easily be  calculated by picking up the term quadratic 
in the imaginary colored chemical potential $A^{a}_{4}$. The result is
   \be
   {\rm Im}\W_{11}^R &=& \frac{1}{4\pi} \left({N_c}\sum_q e_q^2\right)
q^2
     \left[ \left\langle \frac{\alpha_s}{\pi} A_4^2\right\rangle
      \left(\frac{4\pi^2}{T |\vec q|}\right)
      \left( \vphantom{\frac 12} n_+(1-n_+)- n_- (1-n_-) \right)\right]
\ .
   \label{eq:wRnp}
   \ee
According to the estimates in the previous section, 
$\beta\equiv \langle \frac{\alpha_s}{\pi} A_4^2\rangle/T^2$ 
is of the order of $0.4$. Note that (\ref{eq:wRnp})
has the same sign as (\ref{eq:WR}) and will thus increase the 
dilepton yield. In the next section, we will see that this effect can 
be rather dramatic in the invariant mass region studied in the CERES 
experiments.

The contribution of the chromoelectric and chromomagnetic condensates to
the retarded function can be calculated using the methods 
in~\cite{HZ90}, and the result for $q^2 > 0$ is,
   \be
   {\rm Im} \W_{12}^R &=& \frac{1}{4\pi} \left(N_c\sum_q e_q^2\right)
  \left[
   \left( - \frac 16 \left\langle \frac{\alpha_s}{\pi} E^2 \right\rangle 
          + \frac 13 \left\langle \frac{\alpha_s}{\pi} B^2 \right\rangle
   \right)
     \left(\frac{4\pi^2} {T |\vec q|}\right)
      \left( \vphantom{\frac 12} n_+(1-n_+)- n_- (1-n_-) \right) \right]
   \ \ \ . \label{eq:EB}
   \ee
The temperature dependence of (\ref{eq:EB}) is the same as that of
(\ref{eq:wRnp}) and can be understood from the condition of detailed 
balance.  Note that the emission rate is enhanced by a chromomagnetic
field and depleted by a chromoelectric field as expected\footnote{
Roughly speaking, the $e^+e^-$ pairs are repelled from each other
in the CM frame in the presence of a chromoelectric field, and
deflected in the same direction in the presence of a chromomagnetic
field.}, and also that  all condensate contributions vanish at $T=0$.

\section{Results}
\label{sec:results}

\begin{figure}
\centerline{\epsfig{file=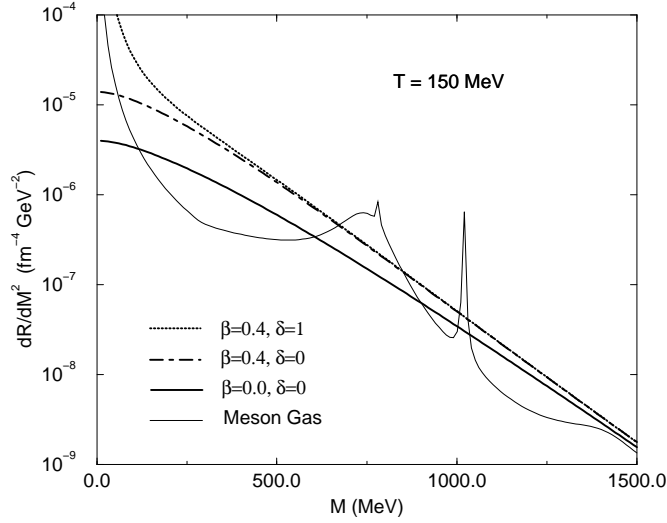,height=8cm}}
\caption{Dielectron rates at T=150 MeV.
$\beta\equiv \langle\frac{\alpha_s}{\pi}A_4^2\rangle/T^2$ and
$\delta \equiv \langle\frac{\alpha_s}{\pi}E^2\rangle/(200\; {\rm MeV})^4
 = \langle\frac{\alpha_s}{\pi}B^2\rangle /(200\; {\rm MeV})^4$.
 See text.}
\label{fig:plasma}
\end{figure}

In Fig.~\ref{fig:plasma} we show the dilepton emission rates at 
temperature $T=150$ MeV, using the parameters, 
$\delta \equiv \langle \frac{\alpha_s}{\pi} E^2\rangle /(200 \; {\rm MeV})^4 =
\langle \frac{\alpha_s}{\pi} B^2\rangle/(200 \; {\rm MeV})^4$ \cite{LAT2}. 
The contribution from the non-perturbative plasma, \ie perturbative plus 
condensate contributions (dotted and dash-dotted lines) should be compared 
to  the purely perturbative part (thick solid line) and, as a reference, 
to that of  a gas of mesons (thin solid line) \cite{LWZ} .  
The condensate contribution to the rates are sizable in the 500 MeV 
region, and the effect of the $\langle A_{4}^{2}\rangle$ condensate, 
$\beta$,  dominates that of the chromoelectric and chromomagnetic ones, 
$\delta$, for masses above 250 MeV. The dependence  on both condensates 
$\beta$ and $\delta$ is negligible at energies higher 
than 2 GeV, and  above 2.5 GeV the 
perturbative plasma gives the same result as that of a hadron gas, as 
expected from duality. In the Dalitz region ($<200$MeV) the effects of 
both the electric and magnetic condensates are appreciable and 
dominates the perturbative part by orders of magnitude. 
This observation maybe important for photon emission as well.

In a heavy-ion collision the electromagnetic emission occurs from
various
stages of the collision process, and  hence at various temperatures. In this
part, 
we will focus on the emission rates in the low and
intermediate mass region (up to 1.5 GeV). We will assume that the 
dileptons  emanate from a simple fire ball composed of the 
nonperturbative plasma phase only.  The fire ball expansion is assumed 
to be homogeneous, and we will use  a standard parametrization of
volume and   temperature,
     \be
     V(t) &=& V_0\left(1+\frac{t}{t_0}\right)^3 \nn
     T(t) &=& (T_i-T_{\infty})e^{-t/\tau}+T_\infty
    \ \ \ . \label{fire} 
    \ee
The parameters in (\ref{fire}) regarding the initial and freeze-out 
temperature will be taken from present transport calculations
\cite{Li,Rapp}. We will not try to justify this point except by saying 
that it allows us to define a space-time volume for the ``hot'' spot in 
the heavy-ion collision that is consistent with the one used for an
interacting meson-phase~\cite{LWZ,Li,Rapp}. Hence $t_0 =10$ fm/c, $T_i = 170$ MeV, 
$T_\infty= 110$ MeV, $\tau=8$ fm/c, and the value of $V_0$ is absorbed 
into the over-all normalization constant $N_0 V_0 =6.76\times 10^{-7}$ fm$^3$.
The freeze-out time will be set to $t_{f.o.}=10$ fm/c. 

\begin{figure}
\centerline{\epsfig{file=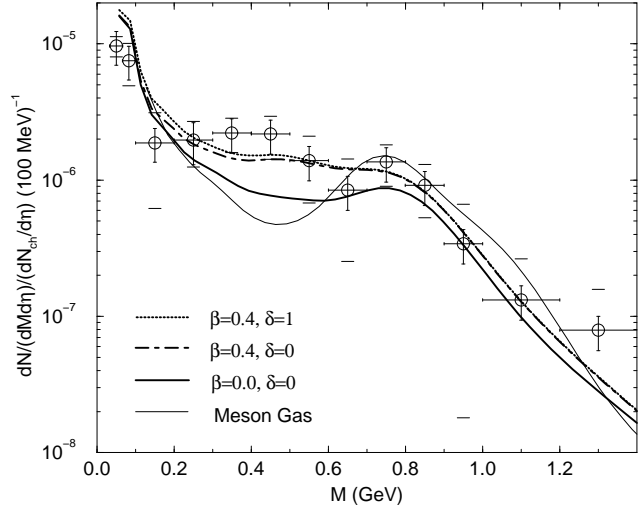,height=8cm}}
\caption{Dielectron rates for CERES S-Au experiment. See text.}
\label{fig:ceres_plasma}
\end{figure}

Using (\ref{fire}) and the above rate, the final expression for the
integrated emission rate per unit rapidity $\eta$ and invariant mass $M$
is
     \be
     \frac{dN/d\eta dM}{dN_{ch}/d\eta} (M)
      &=& N_0 M \int_0^{t_{f.o.}} dt V(t)
     \int\frac{d^3 q}{q_0} A(q_0,q^2)\frac{d\R}{d^4q} \ .
     \label{eq:rate}
     \ee
The acceptance function $A$ enforces the detector cut 
$p_t> 200 MeV, 2.1<\eta<2.65$, and $\Theta_{ee}> 35$ mrad for CERES. 
In Fig.~\ref{fig:ceres_plasma},
we show the evolved emission rates for CERES.
The enhancement is sizable and goes in the direction of the data in the
500
MeV range in CERES. Again, the nonperturbative QCD results are indicated
by
the dotted line for the condensates we quoted, the dashed-dotted line
(without
the E- and B-condensates), while the perturbative QCD result is
indicated by 
the solid line and the meson result by the thin solid line. 

\begin{figure}
\centerline{\epsfig{file=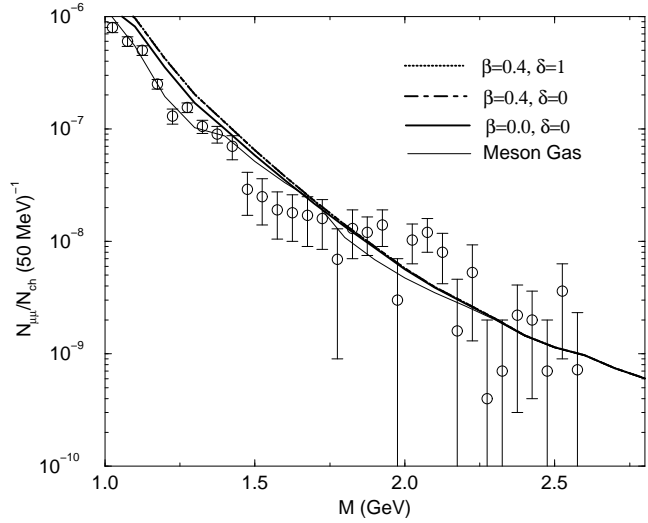,height=8cm}}
\caption{Dimuon rates for HELIOS-3 S-W experiment. See text.}
\label{fig:helios_plasma}
\end{figure}

For HELIOS-3, (\ref{eq:rate}) can be used by integrating over $\eta$
with the cut $m_t\ge 4 (7-2\eta), m_t\ge 
\sqrt{(2 m_\mu)^2+(15/\cosh(\eta))^2}$. In Fig.~\ref{fig:helios_plasma}, 
we show the evolved emission rates for HELIOS. For this energy range the
effect of chromoelectric and chromomagnetic condensates are small irrespective
of the specific value for the thermal condensates as indicated in 
Figs.~\ref{fig:plasma} and \ref{fig:ceres_plasma}.
The enhancement is compatible with the HELIOS data, and
slightly overshoots both the data and the meson gas in the 1 GeV region.

\section{$p_t$ Cuts}
\label{sec:ptcut}

\begin{figure}
\centerline{\epsfig{file=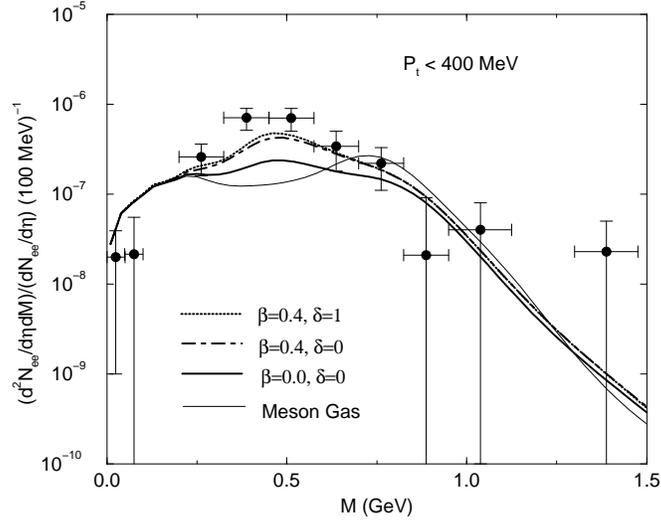,height=8cm}}
\caption{Dielectron rates for low $p_t$ of CERES Pb-Au experiment.}
\label{fig:plasma_pt_low}
\end{figure}

\begin{figure}
\centerline{\epsfig{file=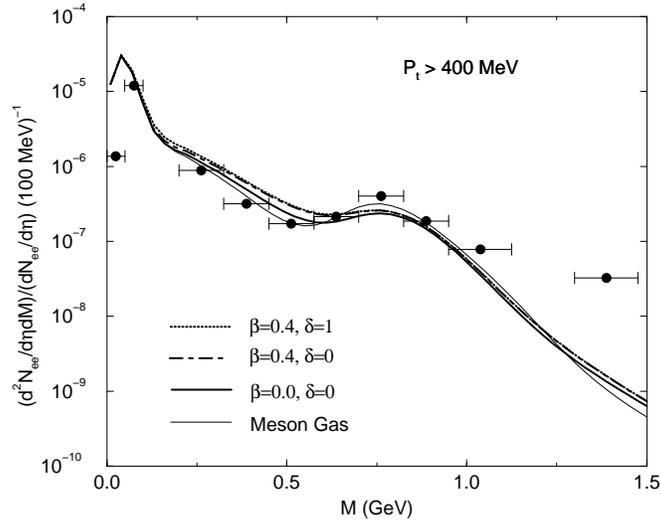,height=8cm}}
\caption{Dielectron rates for high $p_t$ of CERES Pb-Au experiment. 
See text.}
\label{fig:plasma_pt_high}
\end{figure}

\begin{figure}
\centerline{\epsfig{file=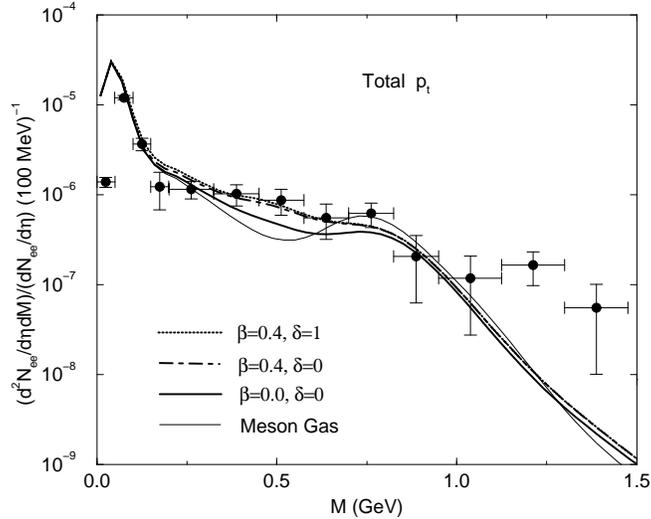,height=8cm}}
\caption{Dielectron rates for the total $p_t$ of
CERES Pb-Au experiment.}
\label{fig:plasma_pt_total}
\end{figure}

Another important check on the nonperturbative emission rates 
can be obtained by using  the recent $p_t$ cut analysis by the CERES 
collaboration to probe the transverse momentum content
of the dilepton emission rates in the fire-ball. Our results versus
the data are shown in 
Figs.~\ref{fig:plasma_pt_low}-\ref{fig:plasma_pt_total}, with and 
without the chromoelectric and chromomagnetic condensates. The
$p_t$ dependent effects caused by the thermal condensates are
most dramatic in the low $p_t$  region (Fig.~\ref{fig:plasma_pt_low}),
where there is a sizable condensate effect, mainly due to 
$\beta$. The condensate
contributions are in the right direction around the low mass 
dielectron enhancement, and give  important additions  to 
the perturbative result.

In the high $p_t$ region, Fig.~\ref{fig:plasma_pt_high}, the
resulting condensate effects are rather small, and the
curve is consistent  with the present data. Therefore,
the total transverse momentum content (Fig.~\ref{fig:plasma_pt_total}) 
is quite accurately reproduced by the addition of the gluon condensates. 

In summary, we find it encouraging that not only 
the dilepton emission rates, but also the observed
low mass-region $p_t$ enhancement 
is, at least partially, explained by our simple model, and 
we think that this further strengthens our proposal
that  nonperturbative condensate effects  play a role in 
the plasma phase of heavy ion collisions. 

%
%
%

\section*{Acknowledgements}

We thank a number of our colleagues at Stony-Brook for discussions.
We also thank Keijo Kajantie and Su Houng Lee for their comments on
the manuscript. This work was supported in part by the U.S. Department 
of Energy under Grant No. DE-FG02-88ER40388, and the Swedish Natural 
Science Research Council.

\end{document}